\documentclass[twocolumn,english,aps,pra,nofootinbib]{revtex4}

\usepackage[T1]{fontenc}
\usepackage[latin9]{inputenc}
\usepackage{amsmath}
\usepackage{amssymb}
\usepackage{stmaryrd}
\usepackage{graphicx}

\makeatletter
\@ifundefined{textcolor}{}
{%
 \definecolor{BLACK}{gray}{0}
 \definecolor{WHITE}{gray}{1}
 \definecolor{RED}{rgb}{1,0,0}
 \definecolor{GREEN}{rgb}{0,1,0}
 \definecolor{BLUE}{rgb}{0,0,1}
 \definecolor{CYAN}{cmyk}{1,0,0,0}
 \definecolor{MAGENTA}{cmyk}{0,1,0,0}
 \definecolor{YELLOW}{cmyk}{0,0,1,0}
}


\makeatother

\usepackage{babel}
\begin{document}

\title{Quantum speed limit helps interpret geometric measure of entanglement}

\author{\L ukasz Rudnicki}
\email{lukasz.rudnicki@ug.edu.pl}

\affiliation{International Centre for Theory of Quantum Technologies (ICTQT),
University of Gda\'{n}sk, 80-308 Gda\'{n}sk, Poland}

\affiliation{Center for Theoretical Physics, Polish Academy of Sciences, Aleja
Lotnik{\'o}w 32/46, PL-02-668 Warsaw, Poland}

\begin{abstract}
Using the approach offered by quantum speed limit, we show that geometric measure of multipartite entanglement for pure states [Phys. Rev. A \textbf{68}, 042307(2003)] can be interpreted as the minimal time necessary to unitarily evolve a given quantum state to a separable one. 
\end{abstract}
\maketitle
\section{Introduction}\label{Sec1}

More than a decade ago, rigorous quantification of entanglement was
in the core of quantum information research \cite{Horo}. Later on, a similar
effort has been directed towards a more general theory of quantum
resources \cite{Gour}, such as, for example, quantum coherence \cite{Plenio}. 

A general methodology behind quantification of resources like entanglement
and coherence is well established. One needs to define a set of resourceless
states, as well as an accompanying set of free (resourceless) operations
(such which cannot generate a given resource). Then, one is in position
to consider measures of the resource in question. Most importantly,
such non-negative measures must be:
\begin{itemize}
\item faithful, which means they vanish only for resourceless states,
\item monotonic, which means they must not increase under resourceless operations.
\end{itemize}
In addition, accompanying requirements such as additivity, convexity, continuity,
etc. can be imposed on a candidate measure \cite{Plenio2}.

Resource-theoretic approach is a handy way of working with quantum phenomena, as it uses mathematical rigor to help decide about optimal
scenarios and protocols. However, from a physical perspective, there
is yet another piece of information which, \emph{a priori}, is not
offered by the resource theories. Given a valid measure of a resource,
say an entanglement measure, we may ask about an operational interpretation
of a value it can assume. If a given measure outputs the values $0.8$
and $0.6$, while evaluated on two concrete states respectively, we
know that the first state is a more resourceful one. But what is the
physical content of these exemplary values? Can they be expressed in terms of quantities
which possess a clear experimental meaning? Answers to these questions,
as not being by default provided by resource theories, if possible,
need to be supplemented by additional considerations. Note that we are not concerned here with the problem, whether an entanglement measure can be read out from outcomes of a tailored experiment.

Within entanglement theory, there are a few known measures which are
equipped with some operational interpretation \cite{Horo}. Among them, we could distinguish:
\begin{itemize}
\item Distillable entanglement \cite{Plenio2} which, using a different wording than is
done usually, represents the success rate while transforming many
copies of a given entangled state into maximally entangled states.
The exemplary value $0.8$ means that asymptotically (very many copies
of the state are used), one can get $8$ maximally entangled states
by using each $10$ given input states;
\item Entanglement cost \cite{Plenio2} which has a dual interpretation with respect to
distillable entanglement;
\item Robustness of entanglement \cite{Vidal} which tells us ``how much'' of a separable
state we need to add to our state (taking a convex combination of
both), in order to make the final state separable;
\item Various distance measures, which tell us how far away the state is from the
set of separable states.
\end{itemize}
All the above examples enjoy an interpretation which is fully satisfactory
from both information-theoretic and probabilistic point of view. However,
in terms of a common-sense meaning of words, certain interpretations
become elusive. In particular, the distance between an entangled quantum
state and the set of separable states is a mathematical distance which
has little to do with distances usually measured in experiments. We
speak here about a distance in a matrix space, not in our physical
spacetime. Obviously, the value of some distance between two states,
equal to $0.8$ (in which units?), does not correspond to any physical distance measured
in meters. 

This conceptual issue becomes even more visible if we look at the
interpretation of probably the most important entanglement measure \cite{Horo},
namely, the entanglement of formation \cite{Formation}. We can say that this measure
quantifies ``\emph{minimal possible average entanglement over all
pure state decompositions}'' \cite{Plenio2} of the given state. It is
quite hard to imagine a device which in some experimental procedure
(even a thought experiment) at the end outputs a number (in physical
units) which will give a direct meaning to the value $0.8$ (or any
other) of this measure.

We can see that even though giving an interpretation to entanglement
measures (and other resource-theoretic measures) is usually possible
to some extent, it is quite hard to provide an interpretation with
an appealing physical background. Therefore, the aim of this paper
is to show that a particular distance measure \textemdash{} geometric
measure of entanglement \cite{Wei} \textendash{} does have such an interpretation. It turns out that for the case of entangled multipartite pure states this distance can directly
be linked with the \emph{minimal time} necessary to make the state
separable by means of a unitary evolution. Knowing an optimal Hamiltonian responsible for the evolution,
which is always of a qubit type \cite{Campaioli}, this minimal time becomes a simple
function of the entanglement measure, with time scale given by the
energy gap $\delta E$ of the optimal Hamiltonian. For example, if
$\delta E/\hbar=2\:\textrm{GHz}$, which lays within a typical regime
relevant for platforms suitable for quantum computation \cite{Qcomp}, the
value $0.8$ assumed by the geometric measure of entanglement means
that the given state will, during the fastest possible evolution, reach
some separable state in $1.11\:\textrm{ns}$. An another state, described
by the geometric measure of entanglement equal to $0.6$ will be faster,
reaching the set of separable states in $0.89\:\textrm{ns}$. 

The above operational meaning can be considered as physically appealing,
as it converts a distance in an abstract mathematical space, to time,
measured by sufficiently accurate clocks. This conceptual transformation
is possible, due to links with, so called, quantum speed limit \cite{Fleming,Aharonov,Zych}.

The paper is organized as follows. In Sec. \ref{Sec2} we briefly
introduce geometric measure of entanglement and quantum speed limit
for pure states. In Sec. \ref{Sec3} we connect both notions expressing,
the former one by the latter one. Crucially, even though quantum speed
limit only offers a bound on the minimal time, it is always possible
to saturate the inequality \cite{Toffoli} giving an exact relation between both quantities
involved. In Sec. \ref{Sec4} we discuss the case of mixed states,
as well as point towards more general  considerations.

\section{Geometric measure of entanglement and quantum speed limit}\label{Sec2}

We conduct our discussion restricting ourselves to pure states. Usually,
such a choice is dictated by simplicity, however, in this particular
problem it has a more fundamental reason. We go back to this point
in Sec. 4.

Given an arbitrary composite system (in general multipartite) with
$K\geq2$ subsystems, we consider entangled states $\left|\psi\right\rangle $
belonging to the Hilbert space $\mathcal{H}=\mathcal{H}_{1}\varotimes\ldots\varotimes\mathcal{H}_{K}$.
We neither need to specify dimensions of each subsystem, nor assume
that all subsystems are identical.

The family of geometric measures of entanglement for pure states is
defined as \cite{Wei}
\begin{equation}
E_{m}\left[\psi\right]=1-\left(\max_{\phi\in S_{m}}\left|\left\langle \psi\left|\phi\right\rangle \right.\!\right|\right)^{2},\quad m=2,\ldots,K.\label{GEOM}
\end{equation}
By $S_{m}$ we denote the set of $m$-separable pure states \cite{Rudnicki2014}. Therefore
standard, also called completely, separable states are in $S_{K}$,
while $S_{1}=\mathcal{H}$ is the entire Hilbert space. From obvious
reasons $E_{1}\left[\psi\right]\equiv0$, so we omit this trivial
case starting the hierarchy at $m=2$. Any $m$-separable state is
a tensor product of $m$ states. When $m=K$ each state in the product
belongs to a particular Hilbert subspace, while for $m<K$ some states
constituting $\left|\psi\right\rangle $ must belong to more than
one subspace. For example, if $K=3$, the state $\left|\phi_{\textrm{sep}}\right\rangle $=$\left|\phi_{1}\right\rangle \varotimes\left|\phi_{2}\right\rangle \varotimes\left|\phi_{3}\right\rangle $
is completely separable, while the state $\left|\phi_{\textrm{bis}}\right\rangle $=$\left|\phi_{12}\right\rangle \varotimes\left|\phi_{3}\right\rangle $,
with $\left|\phi_{12}\right\rangle \in\mathcal{H}_{1}\varotimes\mathcal{H}_{2}$,
is ``just'' bi-separable. The property that the hierarchy $E_{m}$
covers the whole entanglement landscape of a multipartite scenario
is one of its major advantages, while computational difficulties relevant
for more complex systems are a disadvantage shared with virtually
all other measures. Note that certain paths have been followed in order to allow for experimental usefulness of the idea behind the geometric measure of entanglement \cite{Colle1, Colle2, Colle3}.  

The discovery of quantum speed limit \cite{Mandelstam1991} dates back much before
the theory of quantum entanglement has been developed. Regardless
of that fact, there is a lot of ongoing research in this field, for
example \cite{PhysRevLett.120.070401,PhysRevLett.120.070402,PhysRevLett.110.050402, PhysRevA.103.022210,PhysRevResearch.2.023125}. There is a plethora of scenarios in which quantum
speed limit can be considered, rendering better or worse performance,
depending on the context \cite{PhysRevLett.120.060409,Campaioli2019tightrobust}. We go back to this issue in Sec. 4. For
our purpose we resort here to the most standard variant of quantum speed
limit.

We know, that for two pure states $\left|\psi\right\rangle $ and
$\left|\phi\right\rangle $, and a time evolution governed by a time-independent
Hamiltonian $H$, the time $\tau$ necessary to unitarily evolve $\left|\psi\right\rangle $
into $\left|\phi\right\rangle $ is bounded
\begin{equation}
\tau\geq\hbar\frac{\arccos\left(\left|\left\langle \psi\left|\phi\right\rangle \right.\!\right|\right)}{\Delta H}.\label{QSL}
\end{equation}
The usual standard deviation 
\begin{equation}
\Delta H=\sqrt{\left\langle \psi\right|H^{2}\left|\psi\right\rangle -\left(\left\langle \psi\right|H\left|\psi\right\rangle \right)^{2}},
\end{equation}
does only depend on the initial state $\left|\psi\right\rangle $,
and is obviously time independent. The bound in (\ref{QSL}) can always
be saturated with the appropriate ($\left|\phi\right\rangle $-dependent)
choice of $H$. To simplify the discussion, without loss of
generality, we assume that $\left\langle \psi\left|\phi\right\rangle \right.$
is real and non-negative. This is not a restriction since we work
in a complex projective space, so that all states $e^{i\vartheta}\left|\phi\right\rangle $
are equivalent. The optimal Hamiltonian, denoted by $H_{\textrm{opt}}$,
is known \cite{Campaioli} to be proportional to the $\sigma_{y}$ Pauli matrix\footnote{Note that on page 18 in \cite{Campaioli}, Eq. 2.11 suggests that $H_{\textrm{opt}}$ is proportional to $\sigma_x$, which is just a minor mistake \cite{private}.} 
in the two dimensional subspace spanned by $\left\{ \left|\psi\right\rangle ,\left|\phi\right\rangle \right\} $.
Note that both states do not need to be orthogonal, so one introduces
\begin{equation}
\left|\bar{\psi}\right\rangle =\frac{\left|\phi\right\rangle -\left\langle \psi\left|\phi\right\rangle \right.\left|\psi\right\rangle }{\sqrt{1-\left|\left\langle \psi\left|\phi\right\rangle \right.\!\right|^{2}}},
\end{equation}
so that $\left\langle \psi\left|\bar{\psi}\right\rangle \right.=0$.
The Hamiltonian reads
\begin{equation}
H_{\textrm{opt}}=-i\hbar\omega\left(\left|\psi\right\rangle \left\langle \bar{\psi}\right|-\left|\bar{\psi}\right\rangle \left\langle \psi\right|\right),
\end{equation}
where the frequency $\omega$ gives an energy scale. This energy scale
is not subject to further optimization, as its role is just to set
a time scale. By a simple calculation we can confirm that the state
$e^{-iH_{\textrm{opt}}t/\hbar}\left|\psi\right\rangle $ equals $\left|\phi\right\rangle $
for $t=\tau\left[\psi,\phi\right]$, where
\begin{equation}
\tau\left[\psi,\phi\right]=\frac{\arccos\left(\left|\left\langle \psi\left|\phi\right\rangle \right.\!\right|\right)}{\omega}.
\end{equation}
Additionally, $\left\langle \psi\right|H_{\textrm{opt}}^{2}\left|\psi\right\rangle = \hbar^2\omega^2$ and $\left\langle \psi\right|H_{\textrm{opt}}\left|\psi\right\rangle = 0$, so that  $\Delta H_{\textrm{opt}}= \hbar \omega$, and consequently the inequality (\ref{QSL}) is saturated.

\section{Main result}\label{Sec3}

As before let us consider a composite system $\mathcal{H}=\mathcal{H}_{1}\varotimes\ldots\varotimes\mathcal{H}_{K}$,
and an entangled state $\left|\psi\right\rangle \in\mathcal{H}$. We
ask the question: how fast can the state $\left|\psi\right\rangle $
become $m$-separable with the help of the unitary evolution given
by a global Hamiltonian? Obviously, if $\left|\psi\right\rangle $
already is $m$-separable, the time necessary to achieve that task
will trivially vanish. However, if the state is not $m$-separable,
the time, which we denote as $\tau_{m}\left[\psi\right]$, will be
positive. 

From the previous section we know that for a given $\left|\phi\right\rangle $,
saturation of quantum speed limit (\ref{QSL}) occurs for $H=H_{\textrm{opt}}$.
Therefore 
\begin{equation}
\tau_{m}\left[\psi\right]=\min_{\phi\in S_{m}}\tau\left[\psi,\phi\right]\equiv\frac{1}{\omega}\min_{\phi\in S_{m}}\arccos\left(\left|\left\langle \psi\left|\phi\right\rangle \right.\!\right|\right).\label{Intermediate}
\end{equation}
Moreover, since the arccos function is decreasing we observe that
\begin{eqnarray}
\min_{\phi\in S_{m}}\arccos\left(\left|\left\langle \psi\left|\phi\right\rangle \right.\!\right|\right) & = & \arccos\left(\max_{\phi\in S_{m}}\left|\left\langle \psi\left|\phi\right\rangle \right.\!\right|\right)\nonumber \\
 & = & \arccos\left(\sqrt{1-E_{m}\left[\psi\right]}\right)\nonumber \\
 & = & \arcsin\left(\sqrt{E_{m}\left[\psi\right]}\right).
\end{eqnarray}
Note that the minimum in (\ref{Intermediate}) has been turned into
the maximum present in (\ref{GEOM}). We therefore obtain
\begin{equation}
\tau_{m}\left[\psi\right]=\frac{\arcsin\left(\sqrt{E_{m}\left[\psi\right]}\right)}{\omega},\label{8}
\end{equation}
which is the final result of this derivation. As expected, we can
see that if the state $\left|\psi\right\rangle $ is $m$-separable,
the minimal time is $0$, otherwise it is strictly positive. We also
transfer the ordering among classes of entangled states (provided
by the hierarchy $E_{m}$) to the ordering among the ``speeds''
relevant for the quantum speed limit.

For completeness, after a simple rearrangement we can express the
geometric measures of entanglement in terms of the minimal time
\begin{equation}
E_{m}\left[\psi\right]=\sin^{2}\left(\omega\tau_{m}\left[\psi\right]\right).\label{Final}
\end{equation}

\section{Discussion}\label{Sec4}

We derived the relation (\ref{Final}) which links the geometric measure
of entanglement with the minimal time of unitary evolution. As a by-product
of this analysis, maximization with respect to $m$-separable states
has been replaced by minimization with respect to global Hamiltonians
responsible for the time evolution. This replacement is formal, since
to know the optimal Hamiltonian we need to know the optimal $m$-separable
state and \emph{vice versa}. This implies that if one tries to compute
the minimal time $\tau_{m}\left[\psi\right]$, the effort to be taken
is the same as while computing the sole $E_{m}\left[\psi\right]$.
Still, the gain is on the interpretation side. While the value $E_{m}\left[\psi\right]$
can only be understood in information-theoretic context, $\tau_{m}\left[\psi\right]$
has a crystal clear physical meaning. Using qubit type Hamiltonians
with energy gap 
\begin{equation}
\delta E=2\hbar\omega,
\end{equation}
the multipartite entangled state $\left|\psi\right\rangle $ cannot
be made $m$-separable faster, than is specified by Eq. \ref{8}.
Of course the time needs to be given in relation to the energy scale
$\delta E$, since by increasing the energy we always increase the speed
of time evolution.

\begin{figure}

\begin{centering}
\includegraphics[scale=0.34]{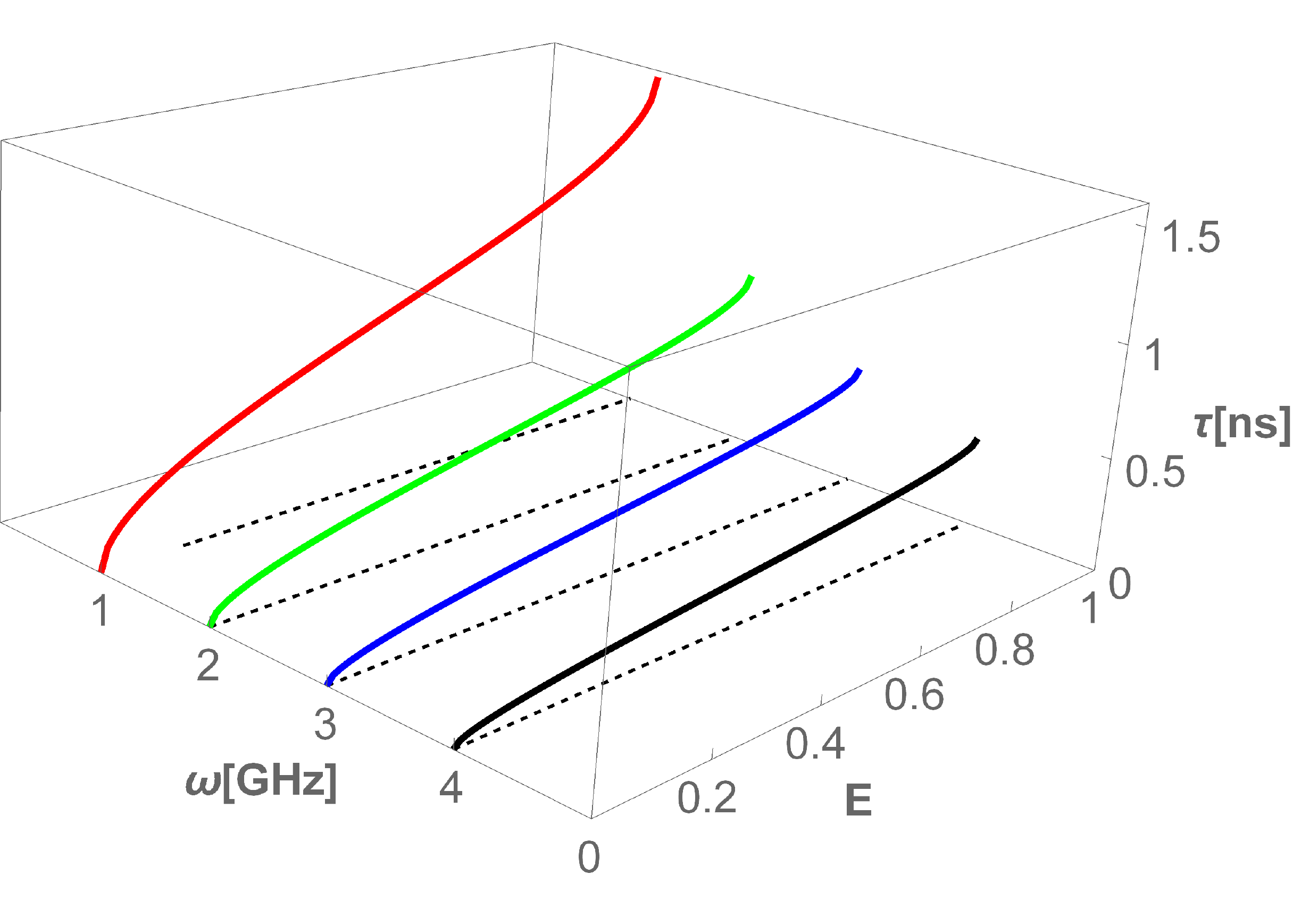}\caption{Minimal time necessary to evolve a given pure state to the closest
separable state. For brevity, the label ``$m$'' and the argument
$\psi$ have been dropped. \label{Fig} }
\par\end{centering}
\end{figure}

In Fig. \ref{Fig}, we plot the minimal time in Eq. \ref{8} for a
few distinct values of $\omega$. We observe that the result does
not explicitly depend on the index ``$m$''. However, we know that
$E_{m}\left[\psi\right]\geq E_{n}\left[\psi\right]$ if $m\geq n$,
so the minimal time $\tau_{m}\left[\psi\right]$ is an increasing
function of $m$.

We are now in position to address the issue of mixed states. Known
entanglement measures do sometimes cover mixed states in a natural
fashion. In majority of cases, however, there are issues with monotonicity,
and one needs to resort to the convex roof construction. Formally
speaking, this construction works always. One needs to represent a
given mixed state as a convex combination of pure states
\begin{equation}
\rho=\sum_{k}p_{k}\left|\psi_{k}\right\rangle \left\langle \psi_{k}\right|,
\end{equation}
compute the measure individually for each $\left|\psi_{k}\right\rangle $,
average over the probability distribution $\left\{ p_{k}\right\} $
and, at the end, optimize with respect to all convex decompositions
of the state in question. Consequently, using an example of the geometric
measure of entanglement, one gets
\begin{equation}
E_{m}\left[\rho\right]=\min_{\left\{ p_{k},\psi_{k}\right\} }\sum_{k}p_{k}E_{m}\left[\psi_{k}\right].
\end{equation}
This construction, even for very simple systems, proves itself to
be intractable, when it comes to concrete calculations. However, as
already mentioned, on the formal level the measure for mixed states
can be defined. 

Our result based on the quantum speed limit \emph{does not} share
this feature. Before we open this part of discussion, we shall observe
that even the convex roof construction, which refers to pure states, is not compatible with our approach,
as the function $\sin^{2}x$ for $0\leq x\leq\pi/2$ is neither convex
nor concave. Therefore, we abandon the convex roof construction and
ask an another question, namely, whether a result similar to (\ref{Final})
could be derived given a mixed state $\rho$, and some entanglement
measure based on the minimal distance between this state and the set
of $m$-separable mixed states. Even in this generalized setting we
immediately meet an obstacle, because in order to unitarily evolve
one state into the other one, both states need to have the same spectrum.
Most likely, the optimal $m$-separable mixed state would not meet
this criterion, therefore, it would be impossible to unitarily evolve
$\rho$ into it. One could of course consider other models for time
evolution, Markovian or non-Markovian, however, \emph{a priori} there
is no guarantee that a fixed model of time evolution will do the job.
Especially, because all known quantum speed limits involving mixed
states are generally not tight \cite{PhysRevLett.120.060409,Campaioli2019tightrobust}. Because of that, one would
eventually end up with a bound on a given entanglement measure. Due
to lack of optimality, the minimal predicted time will be overestimated,
therefore we would likely have an upper bound. While bounds (especially
lower bounds) are useful from a practical perspective, they weaken
the intepretation of the measure in question. In conclusion, the result
reported in this paper from fundamental reasons does not extend to
the case of mixed states. Regardless of that fact, we believe the interpretation valid for pure states stands on its own. As a matter of fact, an interesting future question is related with purification of mixed states and the interplay between the quantum speed limit and the way in which given quantum correlations can be immersed in a larger Hilbert space. 

Since the research devoted to quantum speed limit shifts from pure theory towards applicable quantum information processing \cite{PhysRevLett.126.180603}, it is not surprising that other quantum resources are also in the focus. For example, so called resource speed limit has recently been discussed \cite{Kavan}. On the other hand, formal relatives of the geometric measure of entanglement are also noticed as potentially handy quantifiers of other resources, such as non-classicality in quantum optics \cite{Marian_2020}.

\acknowledgments 
We acknowledge support by the Foundation for Polish Science (IRAP
project, ICTQT, Contract No. 2018/MAB/5, cofinanced by the EU within
the Smart Growth Operational Programme).

\end{document}